\documentclass[prc,preprint]{revtex4}
\usepackage{graphicx}
\usepackage{subfigure}
\usepackage{amsmath,amssymb}
\usepackage{mathrsfs}
\def\dis{distribution}

\def\pt{p_T}
\def\ppt{$p_T$}

\begin{document}

\title{Centrality and Transverse Momentum Dependencies of Minijets and Hadrons in Au-Au Collisions}
\author{Lilin Zhu$^1$ and Rudolph C. Hwa$^2$}
\affiliation
{$^1$Department of  Physics, Sichuan
University, Chengdu  610064, P.\ R.\ China \\
\bigskip
$^2$Institute of Theoretical Science and Department of
Physics\\ University of Oregon, Eugene, OR 97403-5203, USA}

\begin{abstract} 
In the study of hadron production in Au-Au collisions at RHIC minijets play an important role in generating shower partons in the intermediate \ppt\ region. Momentum degradation of the hard and semihard partons as they traverse the inhomogeneous medium at various azimuthal angles results in a complicated convolution of geometrical, nuclear and dynamical factors that cannot usually be described in a transparent way. In this work a compact formula is found that represents the inclusive distributions of minijets of any parton type at the surface of the medium for any collision centrality. They take into account the contributions from all initiating partons created at any point in the medium. By comparing with the case of no energy loss, a ratio has been determined that  is analogous to the nuclear modification factor for minijets. Phenomenological reality of such distributions is examined by calculating the hadronization of the minijets in the recombination model. Good fits of the data on pion, kaon and proton production throughout the intermediate \ppt\ region have been obtained by adjusting  the parameters controlling the magnitude of the thermal partons and the degradation rates of the semihard partons. The result gives support to the minijet spectra at any centrality on the one hand, and the hadronization procedure used on the other. An important property made manifest in this study is that quarks and gluons must not lose energy in the same way because the partons form mesons and baryons differently by recombination and the momenta of quarks and gluons must be degraded at different rates in order to reproduce the experimental pion and proton spectra.
This is a feature that renders invalid the notion of parton-hadron duality or other hadronization schemes based on similar ideas.

\end{abstract}
\maketitle

\section{Introduction}
The study of particle production in heavy-ion collisions is evolving into a mature field, especially for Au-Au collisions at the Relativistic Heavy-Ion Collider (RHIC) \cite{ia, ba, jas,ka,tt, mn}.  Theoretical models that treat the phenomena also seem to settle into different camps, claiming successes in different domains of validity, with hydrodynamical model for transverse momentum $p_T < 2$ GeV/c \cite{vh, tat, hs} perturbative QCD for $p_T > 8$ GeV/c \cite{pq, pq1, pq2, pq3}, recombination model in the intermediate region \cite{hy,vg, rf, hy1,rf1,hwa} and color glass condensate whenever gluon density is high \cite{cgc,cgc1,cgc2,cgc3}.  
Our objective here is to improve the recombination model in various directions: (a) description of momentum degradation at different rates for quarks and gluons at any centrality, (b) minijet distributions of all parton types at medium surface, and (c) $\pt$ spectra of pion, kaon and proton that can reproduce the data over wide ranges of $\pt$ and centrality.

In broadening the \ppt\ range of what we study we do not invalidate other approaches, since our focus is on the hadronization aspect of the problem. On the low-\ppt\ side we overlap with the hydrodynamical model on fluid flow, which does not address the issue of how quarks turn into hadrons. On the high-\ppt\ side our recombination model is consistent with fragmentation, since our shower partons are derived from the fragmentation functions.  What we do not have is a description of the evolutionary process of the hot and dense medium from early time. Because of that deficiency we have two adjustable parameters on the centrality dependence of the magnitude of the thermal distribution. Azimuthal anisotropy is a problem that we have treated previously \cite{hz,hz1} and will not be addressed here.

  Since we describe all the processes in analytical expressions, our presentation has the advantage of showing the details of relevant quantities explicitly instead of being hidden in codes.  In particular, we have found compact formulas to describe the parton momentum distributions at the surface of the medium at mid-rapidity for any centrality, after the hard and semihard partons have undergone momentum degradation in traversing the medium.   Since minijets  play important roles in our description of hadron production, semihard partons that escape the initial thermalization are crucial ingredients in our formalism.  The inclusive distributions of minijets with and without energy loss can jointly lead to the construction of a quantity analogous to the nuclear modification factor, but here for minijets, thereby offering a direct view of the medium effect on partons.  
  
  The minijets generate shower partons after emerging from the medium surface. Those shower partons recombine with themselves or with thermal partons in various combinations to form hadrons. We shall calculate the \ppt\ spectra of pion, kaon and proton for all centralities. Success in adjusting a small number of parameters to achieve agreement with data over a wide range of \ppt\ and centrality secures the affirmation that the formalism is reliable in describing the production of minijets and hadrons.
  
  This paper is organized as follows. In Sec.\ II we outline the basic framework of recombination showing the place where the distribution of minijets is needed. Sec.\ III is where the inclusive distributions of the minijets of all parton species are obtained, and presented in simple parametrized form. The nuclear modification functions for quark and gluon minijets are exhibited in figures. Hadron spectra are calculated in Sec.\ IV and compared to data in Sec.\ V. Concluding remarks are made in the final section.

\section{Basic Framework of Recombination}
We begin with a brief summary of the main equations that are central to our formulation of the recombination model.  They are collected from Refs. \cite{hy,hy1,hz2}, in which details and other references can be found.  The invariant $p_T$ distributions of meson and baryon, averaged over $\eta$ at midrapidity, are
\begin{eqnarray}
p^0{dN^M\over dp_T}&=&\int {dp_1\over p_1}{dp_2\over p_2} F_{q_1\bar q_2}(p_1,p_2) R_{q_1,\bar q_2}^M(p_1,p_2,p_T) ,  \label{1} \\
p^0{dN^B\over dp_T}&=&\int \left[\prod_{i=1}^3 {dp_i\over p_i} \right] F_{q_1q_2q_3}(p_1,p_2,p_3) {R}_{q_1,q_2,q_3}^B(p_1,p_2,p_3,p_T) ,    \label{2}
\end{eqnarray}
where $p_i$ is the transverse momentum (with the subscript $T$ omitted) of one of the coalescing quarks.  $R^{M,B}$ are the recombination functions (RFs) for mesons and baryons, determined previously \cite{hy1,hwa1}.  The $\phi$ dependence has been averaged over, so $dN^h/p_T dp_T$ should be identified with the experimental $dN/2\pi p_T dp_T$ which is integrated over all $\phi$.  The parton distributions can be partitioned into various components, represented symbolically by 
\begin{eqnarray}
F_{q_1\bar q_2}&=&{\cal TT+TS+SS} ,  \label{3} \\
F_{q_1q_2q_3}&=&{\cal TTT+TTS+TSS+SSS},     \label{4}
\end{eqnarray}
where $\cal T$ and $\cal S$ are the invariant distributions of thermal and shower partons, respectively, at late time just before hadronization.

The thermal parton distribution is
\begin{eqnarray}
{\cal T}(p_1) = p_1{ dN^T_q\over dp_1} = Cp_1e^{-p_1/T},     \label{5}
\end{eqnarray}
where $T$ is the inverse slope parameter that need not be identified with the conventional temperature in a hydro model.  It is shown in Ref. \cite{hz1} that both the pion and proton spectra in the region $1 < p_T <2$ GeV/c can be well described by the ${\cal TT}$ and ${\cal TTT}$ components of $F_{q_1{\bar q}_2}$ and $F_{q_1q_2q_3}$ in Eqs.\ (\ref{1}) and (\ref{2}), using a common $T$ for the thermal partons.  It is an important property of the recombination model that the thermal partons are universal irrespective of the hadrons they form at low $p_T$, where shower partons do not have any significant effect on the $p_T$ distributions.

The shower distribution is
\begin{eqnarray}
{\cal S}(p_2)=\int {dq\over q}\sum_i F_i(q) S_i(p_2/q),  \label{6}
\end{eqnarray}
where $S_i(z)$ is the shower-parton distribution (SPD) in a jet of type $i$ with momentum fraction $z$.  The SPD is determined from the fragmentation function \cite{hy2,hy3}.  $F_i(q)$ is the distribution of parton of type $i$ with momentum $q$ at the medium surface before fragmentation.  It depends on centrality and the opaqueness of the medium, and is the quantity that we shall concentrate on in the next section with more care than before.  The density of shower partons plays a crucial role in determining the hadron spectra at intermediate $p_T$.

Formally, the above equations lay the foundation for the calculation of the hadron distributions.  
In the past the minijet distribution $F_i(q)$ has been studied, but not presented in a way that can easily be retrieved for closer examination. In the next section we shall look for an analytical representation of it as a function of centrality and $\phi$.

\section {Momentum Degradation}
The process of momentum degradation on a semihard parton traversing the medium for any centrality and at any angle $\phi$ has been described in Ref.\ \cite{hy4} in a manageable way that can yield scaling results in agreement with the data on the nuclear modification factor for pion, $R^{\pi}_{AA}(p_T,\phi)$, at various centralities \cite{sa}.  We now upgrade that description with the aim to give a better fit to more accurate data on pion and proton production separately and to find simpler parametric formulas that can directly be applied without going through the geometrical details each time.  

Let us start with the basic equation for the parton distribution $F_i(q, \xi)$ in Eq.\ (\ref{6}) with $\xi$ specifying the dynamical path length (to be discussed below)
\begin{eqnarray}
F_i(q,\xi) = \int dk kf_i(k) G(k,q,\xi)      \label{7}
\end{eqnarray}
where $f_i(k)$ is the parton density in the phase space $kdk$ at the point of creation, and $G(k,q,\xi)$ is the momentum degradation function from $k$ to $q$ \cite{hy5}
\begin{eqnarray}
G(k,q,\xi) = q\delta(q-ke^{-\xi}).     \label{8}
\end{eqnarray}
We have used an exponential form for the degradation with the burden being put on $\xi$ to carry all the information on geometry and dynamics.  The distribution $f_i(k)$ of the initial momentum $k$ has been parametrized in Ref.\ \cite{ds}, so our concern is the distribution $F_i(q,\xi)$ of the momentum $q$ at the medium surface.  Since the dynamical path length $\xi$ depends on the nuclear medium and the azimuthal angle, it is more useful for phenomenological purposes to express $F_i(q)$ in terms of measurable quantities: angle $\phi$ and impact parameter $b$ that can be related to the centrality.  Thus we define
\begin{eqnarray}
\bar F_i(q,\phi,b) =\int d\xi P(\xi,\phi,b)F_i(q,\xi),     \label{9}
\end{eqnarray}
which averages over all $\xi$ with the weighting function $P(\xi,\phi,b)$ being the probability of having $\xi$ at $\phi$ and $b$.  This probability function has been studied in detail in Ref.\ \cite{hy4}.  The main points of the geometrical and dynamical considerations are summarized in Appendix A for easy reference here.

Since $P(\xi,\phi,b)$ is properly normalized, the mean dynamical path length is
\begin{eqnarray}
\bar \xi(\phi,b) = \int d\xi \xi P(\xi,\phi,b)
=\gamma \int dx_0dy_0 \ell(x_0,y_0,\phi,b)Q(x_0,y_0,b).     \label{10}
\end{eqnarray}
where $\ell (x_0,y_0,\phi,b)$ is the geometrical path length weighted by nuclear density defined in Eq.\ (\ref{A1}), and $Q(x_0,y_o,b)$ is the probability of production of a hard  (or semihard) parton at the creation point $(x_0,y_0)$ discussed in Appendix A.  
The parameter $\gamma$ represents the dynamical effect of energy loss per unit length.  
In Ref.\ \cite{hy4} a value for $\gamma$ is found for a generic parton sufficient for the purpose of calculating the pion $R^{\pi}_{AA}(\pt,\phi)$. In this paper we aim to determine both the meson and baryon spectra that depend on the quark and gluon \dis s differently, so we shall distinguish $\gamma_q$ and $\gamma_g$ and use $\gamma_i$ as a generic symbol that replaces $\gamma$ in Eq.\ (\ref{10}). Thus $\bar\xi(\phi,b)$ should be labeled by a subscript $i$, and in accordance to Eq.\ (\ref{10}) is proportional to $\gamma_i$, since $\ell(x_0,y_0,\phi,b)$ and $Q(x_0,y_0,b)$ are properties of the nuclear medium only. $P_i(\xi_i,\phi,b)$ depends on $i$ in a trivial way, as is evident in Eq.\ (A2).
Independent of the numerical values of $\gamma_i$, $\bar \xi_i(\phi,b)$ summarizes the $(\phi,b)$ dependence of $P_i(\xi_i,\phi,b)$.  That is, there is a scaling behavior that can be expressed as
\begin{eqnarray}
P_i(\xi_i,\phi,b) = \psi(z)/\bar \xi_i(\phi,b),     \label{11}
\end{eqnarray}
where $\psi(z)$ is a scaling function in the variable
\begin{eqnarray}
z=\xi_i/\bar \xi_i     \label{12}
\end{eqnarray}
and satisfies
\begin{eqnarray}
\int dz\psi(z) = \int dzz\psi(z) = 1.     \label{13}
\end{eqnarray}
Thus $P_i(\xi_i,\phi,b)$ depends only on $\xi_i$ and $\bar \xi_i(\phi,b)$, not on $\phi$ and $b$ separately.  That property offers a remarkable degree of simplicity in the complex geometrical problem of nuclear collisions.  It means that two collisions at different impact parameters may have $i$-type partons produced at different angles $\phi$ that experience the same mean $\bar \xi_i$ and thus can have the same survival rate.

Both $\bar \xi(\phi,b)$ and $\psi(z)$ have been calculated in Ref.\ \cite{hy4} based on Eq.\ (\ref{A3}).  The results are presented in Appendix B in the form of simple analytic formulas that can well approximate the numerical results.  Substituting Eqs.\ (\ref{7}) and (\ref{8}) into (\ref{9}), and then making use of (\ref{12}) we obtain
\begin{eqnarray}
\bar F_i(q,\phi,b) = \int dz\psi(z)q^2e^{2z\bar \xi_i(\phi,b)}f_i(qe^{z\bar \xi_i(\phi,b)}),     \label{14}
\end{eqnarray}
which is a compact equation that relates the distributions of partons at the medium surface with momentum $q$ to the distributions $f_i(k)$ of partons having momentum $k$ at the point of creation anywhere in the medium.  The $\phi$ and $b$ dependencies in Eq.\ (\ref{14}) have been used to show the capability of this formalism to reproduce the PHENIX data on the nuclear modification factor $R^{\pi}_{AA}(p_T,\phi,b)$ for pions \cite{sa}, using $\gamma = 0.11$.  We now upgrade the treatment by differentiating $\xi_i, i=q,g$.

Since the initial parton distribution $f_i(k)$ decreases rapidly with increasing $k$, it is evident from Eq.\ (\ref{14}) that the small $z$ region of the integrand dominates.  It means that the partons that emerge from the surface are more likely to have had a short path length $\xi_i$, which is $z\bar \xi_i(\phi,b)$.    Hence, from both geometrical and dynamical considerations the semihard partons that get out of the medium to produce shower partons are predominantly created nearer to the surface than in the deep interior.

For applications to problems that involve calculating the $p_T$ spectra of any hadron produced, averaged over $\phi$, we note that the $\phi$ variable occurs only in Eq.\ (\ref{14}) since our hadronization process by recombination is local, as is evident from equations shown in Sec.\ II.  Using $c$ to denote centrality (for example, $c=0.05$ for 0-10\%) instead of the impact parameter $b$, we define the distribution averaged over $\phi$ by
\begin{eqnarray}
\hat{F}_i(q,c) = {1\over 2\pi} \int^{2\pi}_0 d\phi \bar F_i(q,\phi,c).     \label{15}
\end{eqnarray}
Because of the appearance of the factor ${1/2\pi}$ here, our hadronic spectrum at midrapidity is $dN_h/p_Tdp_T$ without the ${1/2\pi}$ factor that is exhibited in experimental figures.

We now go into the details of Eqs.\ (\ref{14}) and (\ref{15}) and attempt to find some simple algebraic representation that can help to circumvent the laborious task of dealing with all the intermediate complications involving $\bar \xi_i(\phi,b)$ and $\psi(z)$ each time we need $\hat{F}_i(q,c)$.  Our first step is to note that the initial parton distribution of $f_i(k)$ is given in Ref.\ \cite{ds} only for central collision at $c=0.05$.  For less central collisions the corresponding $f_i(k,c)$ can be determined  by scaling \cite{fms}
\begin{eqnarray}
f_i(k,c) = {T_{AA}(c)\over T_{AA}(0.05)} f_i(k,c=0.05),     \label{16}
\end{eqnarray}
where $T_{AA}(c)$ is the overlap function, given numerically in Ref.\ \cite{33a} and $f_i(k,c=0.05)$ is
\begin{eqnarray}
f_i(k) = K {C' \over (1 + k/B')^\beta}     \label{17}
\end{eqnarray}
with all the parameters $K, C', B'$ and $\beta$ given in \cite{ds}. 
We shall use only the parameters for Au-Au collisions at $\sqrt {s_{NN}}=200$ GeV. 
Combining the above equations 
 we can calculate $\hat{F}_i(q,c)$ for any parton of type $i$ with momentum $q$ at the medium surface.  
 
 In applying $\hat{F}_i(q,c)$ to the calculation of shower partons in the next section, we shall find that gluons contribute more to pions than quarks, while quarks are more important for the formation of protons than gluons because of the valence structure. In being careful in deriving the quark and gluon \dis s $\hat{F}_{q,g}(q,c)$, we further take note of the difference in the rates of energy loss by quarks and gluons, the former being only about half the latter \cite{pq,pq1}. We implement that difference by requiring $\gamma_q\approx \gamma_g/2$ and the average being around 0.11 as given in Eq.\ (\ref{A3}), found in Ref.\ \cite{hy4}. Thus, we set $\gamma_q=0.07$ and
 $\gamma_g=0.14$ with the average being slightly weighted on the gluon side. Since $\bar\xi_i$ is proportional to $\gamma_i$ according to Eq.\ (\ref{10}), our parametrization for $\bar\xi_i(\phi,c)$ can easily be obtained from Eq.\ (\ref{B1}) by modifying the proportionality constant, i.e., $\bar\xi_i=(\gamma_i/\gamma)\bar\xi$.
 
 To facilitate the usage of $\hat{F}_i(q,c)$  in the future, we present the numerical results here in the form of analytic expressions.  We have found that the Tsallis distribution \cite{ct} can fit the $q$ dependence of the numerical results very well 
\begin{eqnarray}
\hat{F}_i(q,c) = A_i(c) \left(1 + {q\over n_iT_t}\right)^{-n_i}     \label{18}
\end{eqnarray}
where $T_t = 0.32$ GeV and
\begin{eqnarray}
A_g(c) = 1450 e^{-c/0.23}&,& \quad n_g=6.66,          \nonumber  \\ 
A_u(c)  = A_d(c) =450 e^{-c/0.21}&,&  \quad n_u=n_d=5.73,        \nonumber  \\ 
A_{\bar u} (c) = A_{\bar d}(c)  = 115e^{-c/0.21}&,& \quad n_{\bar u}=n_{\bar d} = 6.63,       \nonumber   \\  
A_s(c)  = A_{\bar s}(c) = 63 e^{-c/0.21}&,& \quad n_s = n_{\bar s} = 6.96,     \label{19}
\end{eqnarray}
where we have assumed that $\gamma_s=\gamma_q$. 
The gluon to quark ratio increases  with $c$, or decreases with increasing $N_{\rm part}$ because gluons are more likely to lose more energy in larger medium.  
The parameter $T_t$ is universal; it prescribes the small-$q$ behavior that is exponential.  The exponent $n_i$ depends on parton type and specifies the power-law behavior at large $q$.  We show in Fig.\ 1(a) the $q$ dependence of $\hat {F}_g(q,c)$ and in Fig.\ 1(b) $\hat {F}_i(q,c)$ for parton type $i=q (u$  or $d), \bar q$ and $s$, both  for $c=0.05$.  As $c$ increases, the collision becomes more peripheral and the probability of producing minijets decreases exponentially. 

\begin{figure}[tbph]
\vspace*{-2cm}
\includegraphics[width=1\textwidth]{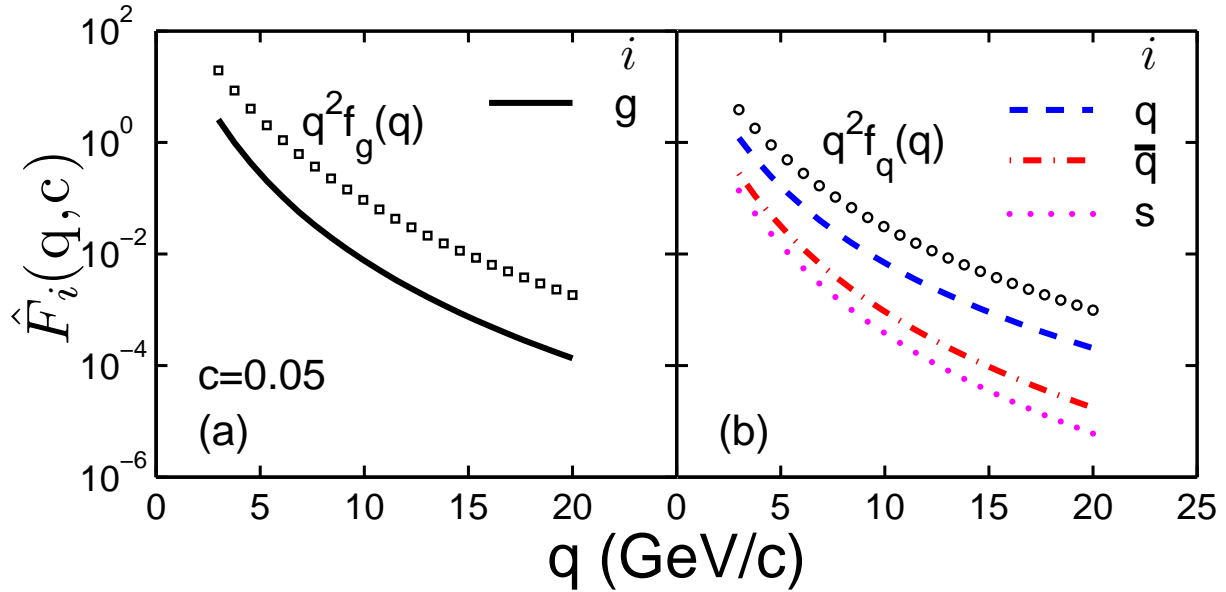}
\vspace*{-2.5cm}
\caption{(Color online) Distributions of minijets at medium surface for 0-10\% centrality. Index $i$ denotes the parton type: (a) $i=g$ for gluon, (b) $i=q, \bar q, s$ (with $\bar s$ being treated the same as $s$). The line with open squares in (a) represents the distribution of gluons without momentum degradation; the line with open circles in (b) represents the same for light quarks.}
\end{figure}

\begin{figure}[tbph]
\vspace*{-.5cm}
\includegraphics[width=1\textwidth]{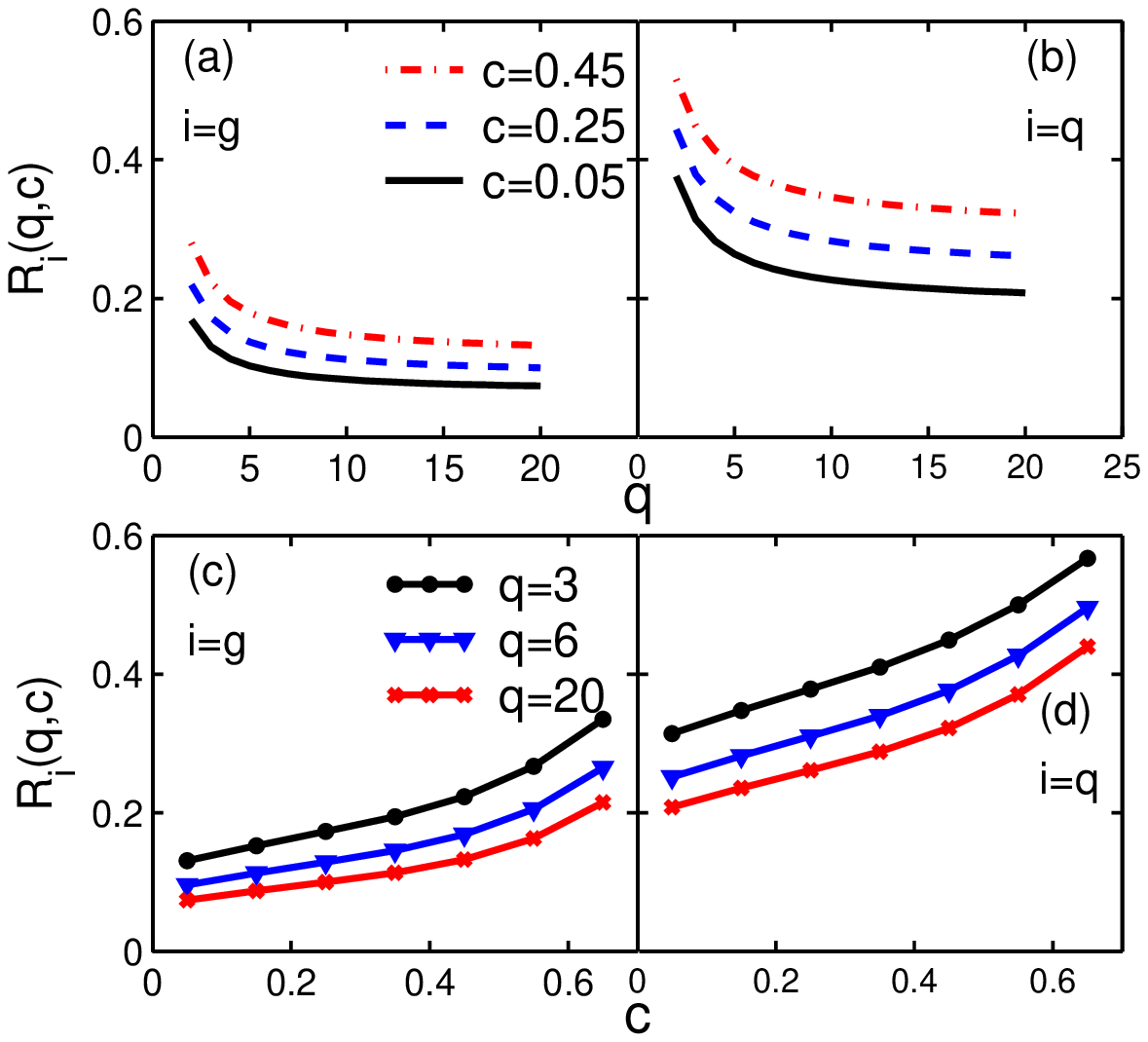}
\caption{(Color online) The ratio $R_i(q,c)$ defined in Eq.\ (\ref{21}) (a) for gluons' dependence on $q$ for fixed centrality $c$, (b) the same as (a) for quarks, (c) dependence on $c$ for gluons at fixed $q$, and (d) the same as (c) for quarks.}
\end{figure}

The decrease of $\hat{F}_i(q,c)$ with increasing $q$ is inherited mostly from $f_i(k)$, but not entirely, as is evident by comparing Eqs.\ (\ref{17}) and (\ref{18}).  From Eqs.\ (\ref{7}) and (\ref{8}) we can determine the relationship between them when there is no momentum degradation, i.\ e., for $c=0.05$,
\begin{eqnarray}
\hat{F}_i(q,0.05) = q^2f_i(q), \quad\quad \xi = 0.     \label{20}
\end{eqnarray}
For gluon, $q^2f_g(q)$ is shown by the line of open symbols in Fig.\ 1(a);
for quarks, $q^2f_q(q)$ is shown similarly in Fig.\ 1(b).
  Comparisons of those two curves of open symbols with the solid (black) line for gluons in (a) and dashed (blue) line for quarks in (b)  show the nature of momentum degradation for $\xi_i \neq 0$.  There is more suppression for gluons than for quarks throughout the  $q$ region shown.  
The degree of degradation can best be revealed by showing the ratio of output ($\xi_i\ne 0$) to input ($\xi_i=0$), i.\ e.\
\begin{eqnarray}
R_i(q,c) = \hat{F}_i(q,c)/[q^2f_i(q,c)],     \label{21}
\end{eqnarray}
where the denominator includes the $c$ dependence due to the nuclear overlap factor exhibited in Eq.\ (\ref{16}).  Thus $R_i(q,c)$ describes the suppression effect due to energy loss, and is 1 if the dynamical path length $\xi_i = 0$.  In Fig.\ 2 we show those ratios for (a) gluons and (b) quarks  as  functions of $q$ and  then in (c) and (d) for their dependencies on $c$. 
$R_i(q,c)$ is analogous to the nuclear modification factor $R_{AA}^h(\pt,c)$, except that it is for $i$-type minijet and is not directly verifiable by experiment. 

In Fig.\ 2(a) and (b) 
we do not go below $q=2$ GeV/c because minijets are ill-defined at lower $q$ and $f_i(k)$ is given for $k>2$ GeV/c in \cite{ds}. 
The suppression of $R_i(q,c)$ is due to a combination of three factors: $\xi_i$ being non-zero,  the necessity for the initial momenta $k$ to be larger than the exit momenta $q$, and the rapid damping of $f_i(k)$ as $k$ increases. 
The increase of $R_i(q,c)$ with decreasing $q$ is more rapid at low $q$. That is because semihard initial partons with $k\sim 10$ GeV/c can be converted to the lower $q$ exit partons by degradation, but hard partons with higher $k$ are rarer and thus ineffective in raising the intermediate $q$ partons. What happens to the energy that is lost is a different issue, and is not a part of $\hat F_i(q,c)$. 
For non-central collisions $R_i(q,c)$ is higher, as one should expect, since there is less suppression when there is less nuclear medium. The $c$ dependencies are shown explicitly in Fig.\ 2(c) and (d) for three typical values of $q$. Comparing Fig.\ 2(a) and (b) we see that $R_g(q,c)$ is roughly half of $R_q(q,c)$, but their $q$ dependencies are similar in shape. The enhanced degradation of gluon momentum is compensated by the higher initial gluon \dis\ $f_g(k)$ compared to  that for quarks $f_q(k)$.  Thus the resultant $\hat F_g(q,c)$ is roughly of the same magnitude as $\hat F_q(q,c)$ at all $q$, as can be seen in Fig.\ 1, by comparing the black solid line in (a) to the blue dashed line in (b). That has a consequence in the production of hadrons at intermediate and high \ppt, as we shall see in the following sections. 

It is important to 
emphasize that $R_i(q,c)$ is not how much momentum fraction that a given parton retains after traversing the medium, but the probability of having a minijet with momentum $q$ relative to no energy loss ($\xi = 0$) after integrating over all contributing sources, all creation points, all azimuthal angles, and especially all initial parton momenta.  
Thus in the hypothetical and unrealistic case of more abundant partons at large $k$, $R_i(q,c)$ would be greater than 1 even if $\xi>0$ because higher $k$ hard partons can feed the lower $q$ minijets after degradation. In reality it is the 
steepness of the falling of $f_i(k)$ at high $k$ that results in the low value of $R_i(q,c)$ even if the degree of degradation is not severe.

Figures 1 and 2 are the results of this study that present new understanding of the properties of minijets produced at intermediate $q$ before hadronization with all geometrical and nuclear complications at all centralities taken into account.  
There are, of course, approximations made in the calculation, most notably in treating energy loss per unit length weighted by nuclear density, i.\ e.\ $\gamma_q = 0.07$ for quarks and $\gamma_g=0.14$ for gluons, which are based on phenomenology done previously. The ultimate test of how good those approximations are is to be found in comparing their consequences to the experimental data that we shall examine in the next two sections.

\section {\large Inclusive Distributions of Hadrons}
Having determined the semihard parton distributions $\hat F_i(q,c)$ for all species and any centrality, we can now return to Sec.\ II and be more explicit about hadron formation by recombination.  Our focus will be on pion, kaon and proton only.  Other mesons and hyperons can be studied in similar ways.  The formalism for recombination of thermal and shower partons has been developed previously \cite{hy,hy1,hz2}.  We generalize to non-central collisions here, and show more explicitly the contributions from various species of semihard partons.  The equations given below can easily be expanded to show $\phi$ dependence if $\hat{F}_i(q,c)$ is replaced by $\bar F_i(q,\phi,c)$, although only the former has been parametrized compactly by Eq.\ (\ref{18}).  

The RFs in Eqs.\ (\ref{1}) and (\ref{2}) are given in Refs.\  \cite{hy1,hz1,hz2}, and will not be repeated here beyond the simplest case for pion
\begin{eqnarray}
R^{\pi}(p_1,p_2,p_T) = {p_1p_2\over p_T} \delta(p_1+p_2-p_T).     \label{23}
\end{eqnarray}
All RFs have the momentum conservation $\delta$-functions.  The prefactors depend on the hadronic wave functions in the momentum space of the constituents \cite{hwa1}.  The thermal parton distribution is shown in Eq.\ (\ref{5}), but the normalization factor $C$ will be given centrality dependence below.  The inverse slope $T$ is independent of centrality because ${\cal T} (p_1)$ is the distribution at the time of hadronization and has the same $p_1$ dependence at any centrality.  We shall assume that the $s$ quarks are equilibrated with the light quarks so $T_s = T$.  The shower distribution ${\cal S}(p_2)$ has the generic form given in Eq.\ (\ref{6}) and will be made more explicit with superscript $j$ to denote the quark type that undergoes recombination.  The cases for pion, kaon and proton are considered separately below.

\subsection {Pion Production}
It follows from Eqs. (\ref{1}), (\ref{3}), (\ref{5}) and (\ref{23}) that for TT recombination we obtain
\begin{eqnarray}
{dN^{TT}_{\pi}\over p_Tdp_T} = {C^2\over 6} e^{-p_T/T}     \label{24}
\end{eqnarray}
where the normalization factor $C$, which has dimension (GeV)$^{-1}$, depends on the number of participants $N_{\rm part}$ as
\begin{eqnarray}
C(N_{\rm part}) = C_0N^{\omega}_{\rm part}.     \label{25}
\end{eqnarray}
In Ref.\ \cite{hz1} $C_0$ and $\omega$ are given different values for $\pi$ and $p$.  We shall find common values for them below.

For TS recombination (and in all other cases where consideration of explicit charge states can promote clarity) let us focus on $\pi^+$ production specifically, although the result is charge independent.  We have two terms ${\cal T}^{\bar d}{\cal S}^u + {\cal T}^u{\cal S}^{\bar d}$, but ${\cal T}^{\bar d} = {\cal T}^u$.  Thus the $\pi^+$ spectrum is
\begin{eqnarray}
{dN^{TS}_{\pi^+}\over p_Tdp_T} = {C\over p^3_T} \int^{p_T}_0 dp_1p_1e^{-p_1/T}[{\cal S}^u(p_T-p_1,c) + {\cal S}^{\bar d}(p_T-p_1,c)],     \label{26}
\end{eqnarray}
where
\begin{eqnarray}
{\cal S}^j(p_2,c) = \int {dq\over q} \sum_i \hat{F}_i(q,c)S^j_i(p_2,q).     \label{27}
\end{eqnarray}
The SPDs $S^j_i$ in a minijet with momentum $q$ have been studied in detail in Ref.\ \cite{hy2}.  We summarize their essential properties in Appendix C.  In the notation discussed in that Appendix, we can exhibit the summation in the integrand in Eq.\ (\ref{27}) more fully, for $j=u$, as
\begin{eqnarray}
\sum_i \hat{F}_iS^u_i = \hat{F}_gG + \hat{F}_uK_{NS} + \left(\sum_{i=i_q,i_s} \hat{F}_i \right) L.     \label{28}
\end{eqnarray}
For $j=\bar d$, only the second term needs to be changed form $\hat{F}_u$ to $\hat{F}_{\bar d}$.  They are the "valence" contributions in the jets.

For SS recombination, which is equivalent to fragmentation, we have
\begin{eqnarray}
{dN^{SS}_{\pi^+}\over p_Tdp_T} = {1\over p_T} \int {dq\over q^2} \sum_i \hat{F}_i(q,c) D^{\pi^+}_i(p_T,q),     \label{29}
\end{eqnarray}
where
\begin{eqnarray}
\sum_i \hat{F}_i D^{\pi^+}_i = {1\over 2} \left[\hat{F}_gD^{\pi^\pm}_g + \sum_{i_q} \hat{F}_{i_q}D^{\pi^\pm}_u + \sum_{i_s} \hat{F}_{i_s}D^{\pi^\pm}_s \right].     \label{30}
\end{eqnarray}
The factor 1/2 is due to the fact that only the FFs $D^{\pi^\pm}_i$ are given for $i \rightarrow \pi^+ + \pi^-$ with $i=g, u$ and $s$ \cite{kkp}.

\subsection {Kaon Production}
While pion mass is neglected above, kaon mass is not negligible, so $p^0$ in Eq.\ (\ref{1}) becomes $m^h_T = (m^2_h + p^2_T)^{1/2}$ in the following.  Difference in the constituent quark masses between $m_q$ and $m_s$ results in asymmetry of the RF for kaon.  Otherwise, the three components of the kaon inclusive distribution are similar to those of the pion.  We simply write them here for $K^+$ production \cite{hy1,hz2,hy6}.
\begin{eqnarray}
{dN^{TT}_{K^+}\over p_Tdp_T} &=& {C^2\over 5} {p_T\over m^K_T} e^{-p_T/T},     \label{31}\\
{dN_{K^+}^{TS}\over p_Tdp_T} &=& {12C\over m_T^Kp_T^5} \int_0^{\pt} dp_1 p_1^2(\pt-p_1)^2  \nonumber  \\
&&\times \left[e^{-p_1/T}{\cal S}^{\bar s}(p_T-p_1,c) + \left({\pt\over p_1}-1\right)e^{-(\pt-p_1)/T}{\cal S}^u(p_1,c)\right],     \label{32} \\
{dN^{SS}_{K^+}\over p_Tdp_T} &=& {1\over m^K_T} \int {dq\over q^2} \sum_i \hat{F}_i(q,c)D^{K^+}_i(p_T,q).     \label{33}
\end{eqnarray}
The shower distribution in Eq.\ (\ref{32}) is as in (\ref{27}), except that for ${\cal S}^{\bar s}$ the summation over $i$ differs from Eq.\ (\ref{28}) as follows
\begin{eqnarray}
\sum_i \hat{F}_i S^{\bar s}_i = \hat{F}_g G_s + \hat {F}_{\bar s}K_{NS} + \left(\sum_{i=i_q,i_s} \hat{F}_i \right)L_s.     \label{34}
\end{eqnarray}
The summation in Eq.\ (\ref{33}) is
\begin{eqnarray}
\sum_i \hat{F}_i D^{K^+}_i = {1\over 2} [\hat{F}_gD^{K^\pm}_g + (\hat{F}_u + \hat{F}_{\bar u} + \hat{F}_s + \hat{F}_{\bar s}) D^{K^\pm}_u + (\hat{F}_d + \hat{F}_{\bar d})D^{K^\pm}_d].     \label{35}
\end{eqnarray}
We note that at low \ppt\ where TT dominates the kaon spectrum differs from the pion spectrum mainly by the $\pt/m_T^K$ factor, while at intermediate \ppt\ the TS components are different not only because of kinematical factors, but also dynamically due to ${\cal S}^{\bar s}$ being more suppressed compared to ${\cal S}^{\bar d}$, as can be seen in
Eq.\ (\ref{19}) as well as in $G_s$ vs $G$ in
 Eq.\ (\ref{28}) and (\ref{34}). Nevertheless, apart from differences in magnitudes, the \ppt\ dependencies are rather similar between $K$ and $\pi$. There are, however, contributions to the pion spectra from resonance decay  at very low \ppt\  that will be considered in the next section.

\subsection {Proton Production}
The RF for proton not only is more complicated due to the three-quark structure but is also known more precisely because of its relation to the proton structure that has been probed exhaustively in deep inelastic scattering.  It is given explicitly in Refs. \cite{hy1,hy6,hy7}, and will be used below with $\alpha = 1.75$ and $\beta = 1.05$.  The result for TTT recombination is then
\begin{eqnarray}
{dN^{TTT}_p\over p_Tdp_T} = g^p_{\rm st}N_pN'_p {C^3p^2_T\over m^p_T}e^{-p_T/T},     \label{36}
\end{eqnarray}
where $g^p_{\rm st} = 1/6$ \cite{hy}, and 
\begin{eqnarray}
N_p &=& [B(\alpha+1, \beta +1)B(\alpha +1, \alpha + \beta +2)]^{-1} ,   \label{37}  \\
N'_p &=& B(\alpha +2, \beta +2)B(\alpha+2, \alpha+\beta+4),     \label{38}
\end{eqnarray}
where $B(\alpha,\beta)$ is Beta function. 
Comparing Eqs.\ (\ref{24}) and (\ref{36}) we note that apart from the common exponential factor, $e^{-p_T/T}$, the thermal-parton contribution to the proton has the additional prefactor $p^2_T/m^p_T$, which is required from simple dimensional consideration:  $C^2$ from TT and $C^3$ from TTT with $C$ having dimension (GeV/c)$^{-1}$ demand another momentum that $p^2_T/m^p_T$ supplies, remembering that $m^p_T$ comes from $p^0$ in Eq.\ (\ref{2}).  Because of this prefactor the proton spectrum deviates from being strictly exponential.

For TTS and TSS we have
\begin{eqnarray}
{dN_p^{TTS}\over \pt d\pt}&=&{g_{st}^pN_p 2C^2\over m_T^p \pt^{2\alpha+\beta+3}} \int_0^{\pt} dp_1 \int_0^{\pt-p_1} dp_2\ e^{-(p_1+p_2)/T}  \nonumber \\
	&& \hspace{1cm} \times (p_1p_2)^{\alpha+1}(\pt-p_1-p_2)^{\beta} {\cal S}^q(\pt-p_1-p_2,c), \label{39} \\
{dN_p^{TSS}\over \pt d\pt}&=&{g_{st}^pN_p 2C\over m_T^p \pt^{2\alpha+\beta+3}} \int_0^{\pt} dp_1 \int_0^{\pt-p_1} dp_2 e^{-p_1/T}  \nonumber  \\
&& \hspace{1cm} \times p_1 (p_1p_2)^{\alpha}(\pt-p_1-p_2)^{\beta}  {\cal S}^{qq}(p_2,\pt-p_1-p_2,c)
\label{40}
\end{eqnarray}
where
\begin{eqnarray}
{\cal S}^{qq}(p_2,p_3,c) = \int {dq\over q} \sum_i {\hat F}_i(q,c)S^q_i(p_2,q)S^q_i(p_3,q-p_2).     \label{41}
\end{eqnarray}
The summation above can be written out more explicitly as
\begin{eqnarray}
\sum_i {\hat F}_iS^q_i(2)S^q_i(3) = {\hat F}_gG(2)G(3) + ({\hat F}_u + {\hat F}_d)K_{NS}(2)L(3) + \left(\sum_{i=i_q,i_s}{\hat F}_i\right)L(2)L(3).     \label{42}
\end{eqnarray}
Finally, for SSS recombination we use FF directly and get
\begin{eqnarray}
{dN^{SSS}_p\over p_Tdp_T} = {1\over m^p_T} \int {dq\over q^2} \sum_i {\hat F}_i(q,c)D^p_i(p_T,q)     \label{43}
\end{eqnarray}
where
\begin{eqnarray}
\sum_i{\hat F}_iD^p_i = {\hat F}_gD^{p/\bar p}_g + {\hat F}_uD^{p/\bar p}_u + {\hat F}_dD^{p/\bar p}_d + \left(\sum_{i=\bar u,\bar d,s,\bar s}{\hat F}_i\right)D_s^{p/\bar p} . \label{44}
\end{eqnarray}

\section {\large Results}
Let us summarize what we have done so far.  In Sec.\ III we have treated the momentum degradation problem and determined the distribution ${\hat F}_i(q,c)$
 of semihard parton $i$ emerging from the medium surface as minijet for any centrality.  
 No adjustable parameters have been used beyond what has previously been parametrized.  
 Equations (\ref{18}) and (\ref{19}) are simple formulas that can well represent the numerical results from detailed calculations based on inputs obtained from earlier studies.  From ${\hat F}_i(q,c)$ we can calculate the shower distributions ${\cal S}^j(p_2,c)$ and ${\cal S}^{qq}(p_2,p_3,c)$ according to Eqs.\ (\ref{27}) and (\ref{41}), using SPDs $S^j_i$ that have previously been determined.  It is then possible to proceed to the calculation of the inclusive distributions of $\pi, K$, and $p$ in Sec.\ IV by including the contributions from thermal partons in various forms of recombination.  We have assumed that the inverse slopes $T$ and $T_s$ are the same and independent of centrality.  
 We take $T$ to be the value $T=0.283$ GeV/c determined in Ref.\ \cite{hz1} without alteration. It is a phenomenological inverse slope that describes the pion and proton spectra at $1<\pt<2$ GeV/c and should not be identified with any temperature in hydrodynamics.
 The 
 only unknown 
 that remains is the $c$ dependence of normalization of the thermal parton, $C(c)$.  We write it in terms of the number of participants as in Eq.\ (\ref{25}) with two undetermined parameters $C_0$ and $\omega$.

It should be noted that the formalism for hadronization described in Sec.\ IV is applied at the final stage of the evolution of the system when the density is low enough for confinement to take place.  Since hadrons are formed by the recombination of quarks (and antiquarks), all gluons have been converted to quark pairs so that no partons are left at the end, although we look at only the single-particle inclusive distributions.  That conversion has been implicitly accounted for in the determination of $S^j_i$ \cite{hy1,hy2} and explicitly in \cite{hy6}, and has been termed saturation of the sea.  The thermal partons have quarks and antiquarks that are fully equilibrated in the light and strange sectors, since $T = T_s = 0.283$ GeV is significantly higher than the $s$ quark mass.  Thus $C(N_{\rm part})$ in Eq.\ (\ref{25}) applies to both sectors.

\begin{figure}[tbph]
\hspace*{-4cm}
\includegraphics[width=1.6\textwidth]{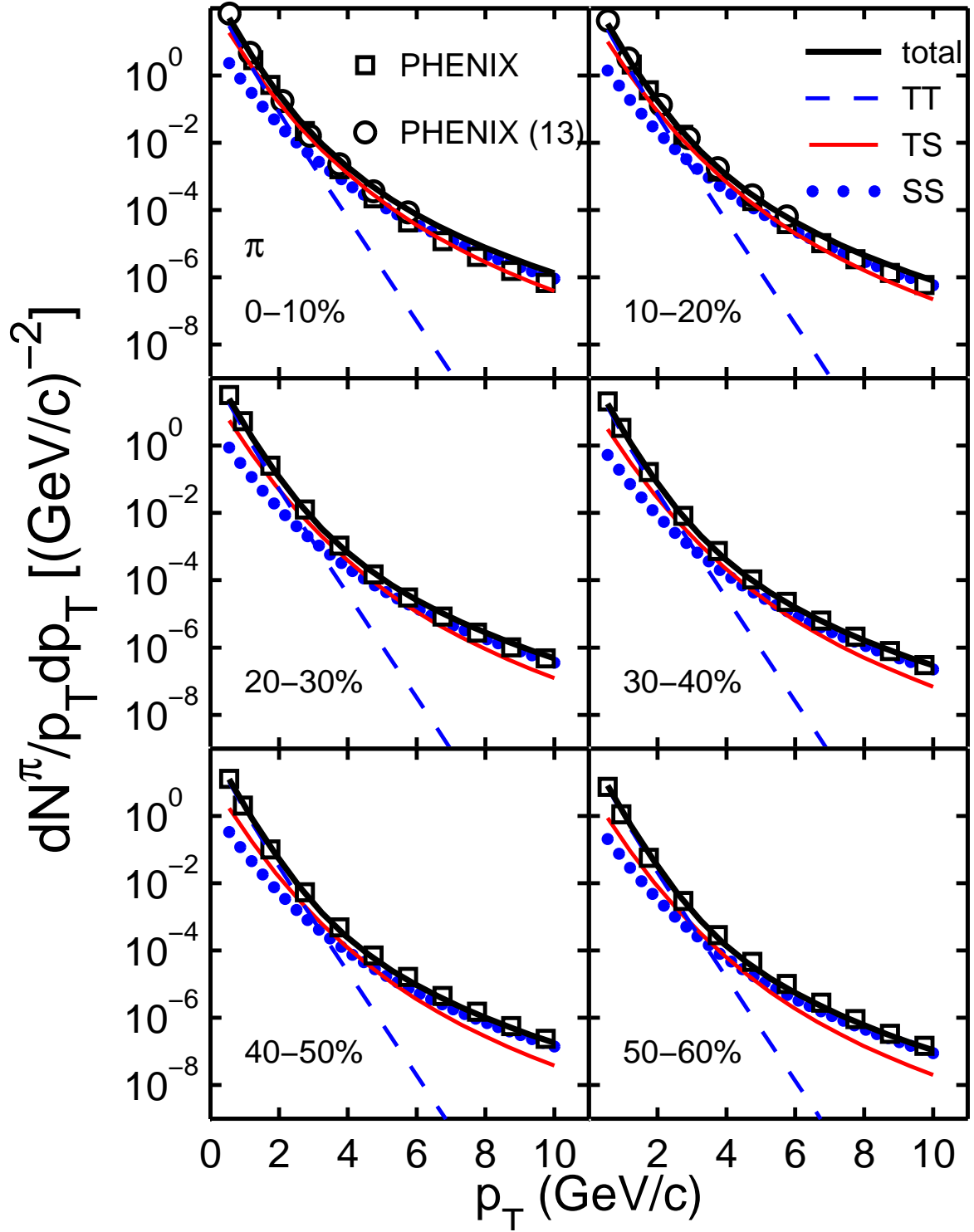}
\caption{(Color online) Pion spectra for 6 centrality bins. The data are from PHENIX: squares \cite{33a,aa}, circles \cite{ad}. }
\end{figure}

\begin{figure}[tbph]
\hspace*{-.5cm}
\includegraphics[width=1.0\textwidth]{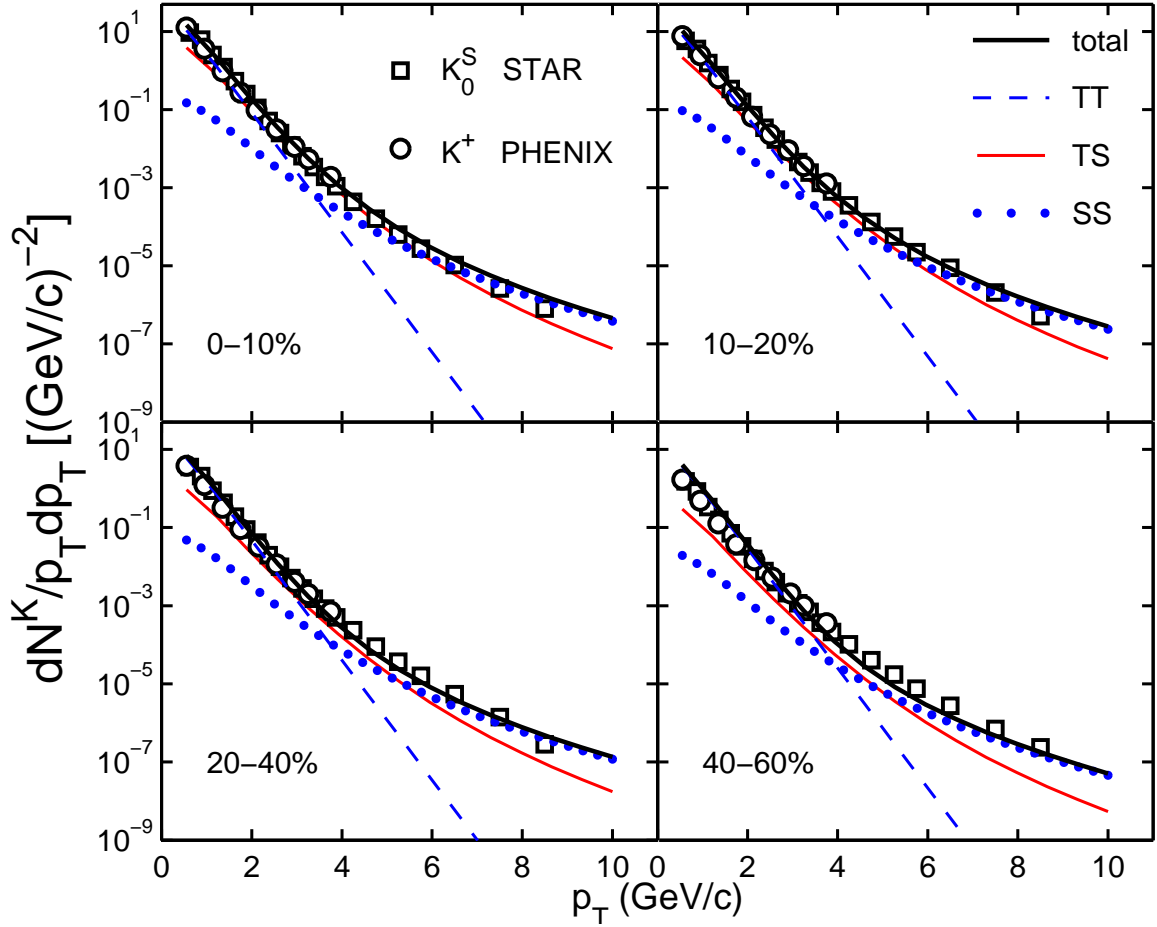}
\caption{(Color online) Kaon spectra for 4 centrality bins. The data are from  STAR: squares \cite{ga}; PHENIX:  circles \cite{ad}.}
\end{figure}

\begin{figure}[tbph]
\vspace*{1cm}
\hspace*{-4cm}
\includegraphics[width=1.5\textwidth]{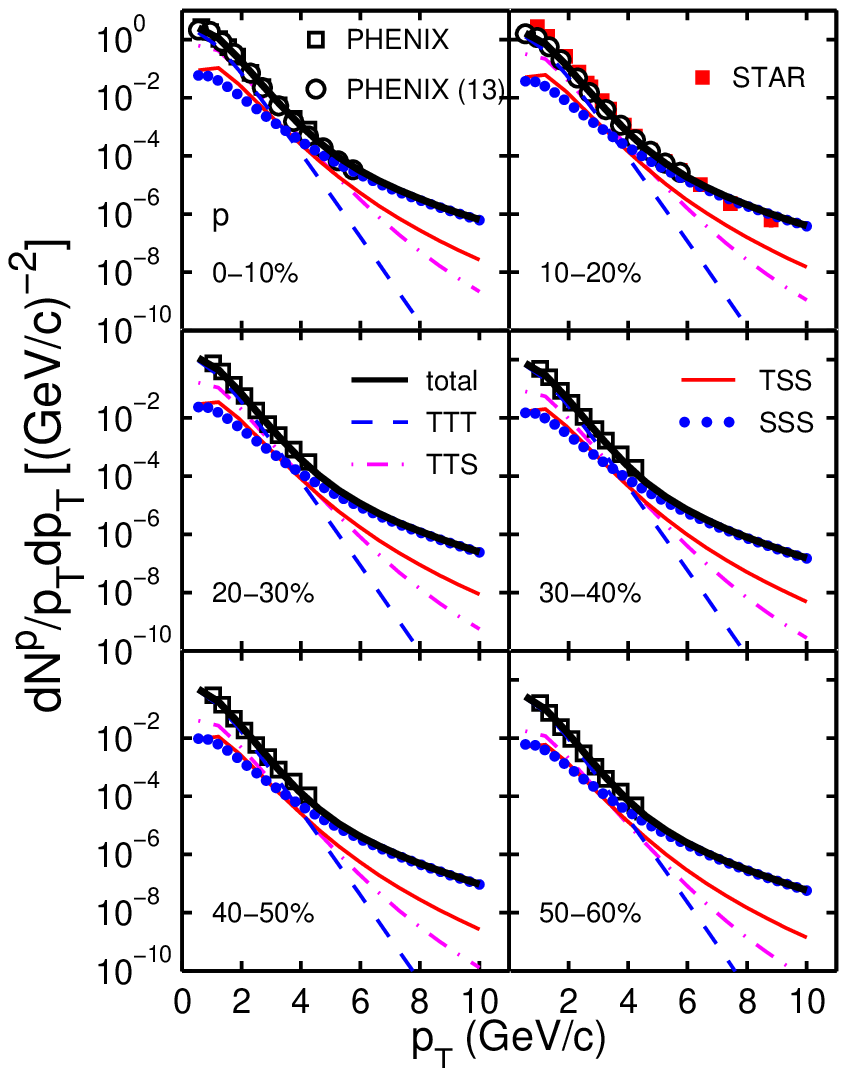}
\caption{(Color online) Proton spectra for 6 centrality bins. The data are from PHENIX: larger (black) squares \cite{33a,ssa}, circles \cite{ad}; STAR: smaller (red) squares \cite{ja}. }
\end{figure}

We are now ready to compute the inclusive distributions of $\pi, K$ and $p$ by varying the two parameters $C_0$ and $\omega$ to fit all the data points.  We emphasize that we have no parameters to adjust for the intermediate and high-$p_T$ regions, since all details on minijets have been specified in Sec.\ III.  In varying $C(N_{\rm part})$ we only adjust the normalization of the thermal distributions.  
In Fig.\ 3 is shown the pion spectra, Fig.\ 4  kaon spectra and Fig.\ 5  proton spectra for various centralities indicated.  The data are from PHENIX and STAR collaborations \cite{33a,ssa,aa,ja,ga,ad}.  The thermal and shower components in various combinations are shown by the different line types.  The dashed lines show the TT and TTT components; their magnitudes are what we have adjusted to fit.  They correspond to the values 
\begin{eqnarray}
C_0 = 3.43, \qquad \omega = 0.32   \label{45}
\end{eqnarray}
in Eq.\ (\ref{25}).  What we find is that while the kaon and proton distributions at low \ppt\ can fit the data well, the pion distribution is a little lower than the data for $\pt<1$ GeV/c. There is a good reason for that, namely: contributions to the pion spectra from the decays of resonances are not included in our calculation of the TT component. Without a definitive scheme to account for resonance decay, we insert a term for very low \ppt\ so that the modified TT component differs from Eq.\ (\ref{24}) as
\begin{eqnarray}
{dN^{TT}_{\pi}\over p_Tdp_T} = [1+u(\pt,N_{\rm part})]{C^2\over 6} e^{-p_T/T}     \label{46}
\end{eqnarray}
where $u(\pt,N_{\rm part})$ is attributed entirely to the effect of resonances
\begin{eqnarray}
u(\pt,N_{\rm part})=(2.8+0.003N_{\rm part}) e^{-\pt/0.65}  \label{47}
\end{eqnarray}
with parameters chosen to fit the pion data at $\pt<1$ GeV/c, approximately the same as in Ref.\ \cite{rhz}.  
The solid lines in Fig.\ 3 are the results that include this term in Eq.\ (\ref{46}). 
We do not regard the presence of this term as a serious weakness of our model; on the contrary, to find an agreement with the data in the absence of it would indicate a problem since resonance production is a reality.

In an overall view of Figs.\ 3-5 it is remarkable how well the solid theoretical curves  fit the data in all 16 cases  by varying just two parameters in the centrality dependence of the thermal distribution.  
In each case the TS, TTS and TSS components play crucial roles in uplifting the spectra in the intermediate region between low $p_T$ where TT and TTT dominate and high $p_T$ where SS and SSS dominate.  That aspect of the $p_T$ behavior has become the hallmark of the success of the recombination model, now extended to all centralities.
It should be noted that the use of $\gamma_q=0.07$ and $\gamma_g=0.14$ is not a free choice of the parameters for parton degradation. We have originally used $\gamma=0.11$ as an approximation for a generic parton determined previously in \cite{hy4}, but were unable to obtain satisfactory result on both pion and proton spectra, slightly high on pion and low on proton at intermediate and large \ppt. We then used $\gamma_q=\gamma_g/2$ with average near 0.11, not arbitrarily, but in recognition of the greater energy loss of gluon compared to quarks by a factor of about 2, as mentioned in Sec.\ III. As a consequence, more quark-type minijets survive the medium effect than the gluons, compared to the case of $\gamma=0.11$, and was sufficient in our treatment of hadronization by recombination to enhance the production of protons and suppress that of pions just enough to render a good fit of the respective spectra. 
It is at this point that we can refer back to the statement made at the end of Sec.\ III and confirm that our treatment of momentum degradation has found support in being able to reproduce the experimental data throughout the whole $p_T$ spectra of produced hadrons.  Thus the minijet distribution ${\hat F}_i(q,c)$ given in Eq.\ (\ref{18}) and (\ref{19}), and shown in Figs.\ 1 and 2, represents reliable information in compact form that can readily be applied in other calculations.

\begin{figure}[tbph]
\hspace*{1cm}
\includegraphics[width=.8\textwidth]{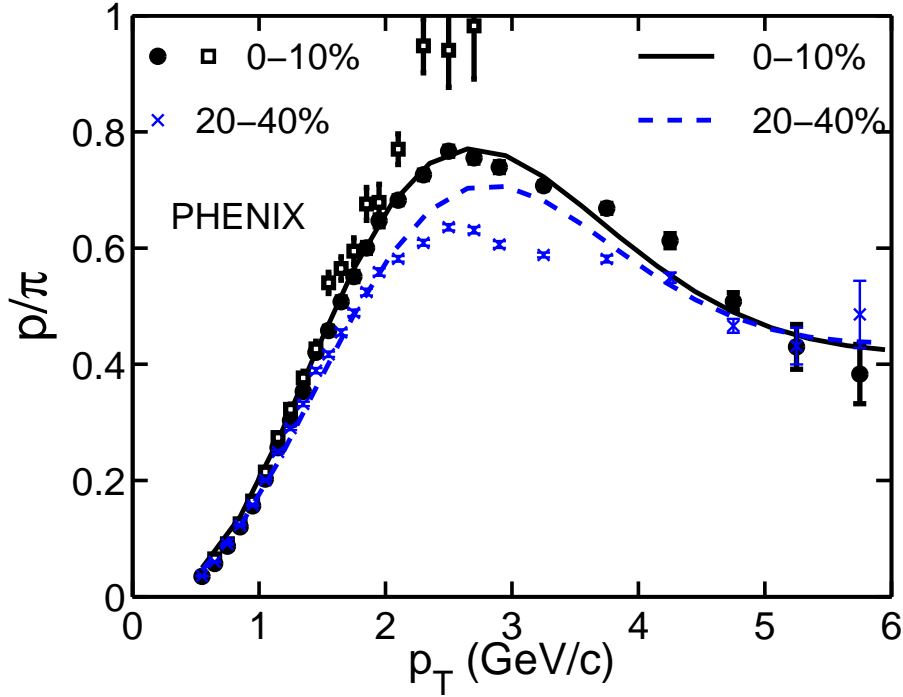}
\caption{(Color online) Proton-to-pion ratio vs $p_T$ for different centralities.  The data are all from PHENIX \cite{ad}, except for the open squares which are from the older data \cite{ssa}.}
\end{figure}

From the spectra obtained we can readily calculate the $p/\pi$ ratio. In Fig.\ 6 we show that ratio for two centralities: 0-10\% and 20-40\%. 
The data from PHENIX \cite{ad} on $p/\pi^+$ show peaking at 
 $\pt\sim 2.5$ GeV/c. 
 Our calculated curve in solid (black) line for 0-10\% agree with the data very well. The dashed (blue) line for 20-40\% peaks at \ppt\ closer to 3 GeV/c, and exceeds the maximum of the data by about 10\%. Nevertheless, the overall shape of the ratio is fairly well reproduced.

\section {\large Concluding Comments}
The main message that this study conveys is that the $p_T$ and centrality dependencies of $\pi, K,$ and $p$ spectra are well reproduced by the recombination model.  Furthermore, our analysis reveals the properties of the partons before hadronization.  Generally speaking, there is some degree of universality in the thermal and shower components that are insensitive to the hadron types, {\it viz.} mesons or baryons, that they form.  On the other hand, there are differences in details that come to light only in the hadronization procedure examined.

The thermal partons have exponential behavior characterized by an inverse slope $T$ that is the same for both light and strange quarks.  Its value at 0.283 GeV is higher than the freeze-out temperature of fluid considered by hydrodynamical studies at the final stage of the medium evolution.  Thus our $T$ is not the hydro temperature, but may be related to an effective temperature that includes the transverse flow energy of the hadrons in some hydro models \cite{ph}.  It is demonstrated in our calculation that a universal $T$ is sufficient to give rise to different low-$p_T$ behaviors of different hadrons because of differences in the recombination functions.  Furthermore, the centrality dependence of the magnitudes of thermal partons is nearly universal.  We caution against being misled by the use of the work ``thermal".  Without having investigated the problem of equilibration, we cannot be certain of the time when thermalization is established.  Our approach allows for minijets to escape from the medium before equilibration, and the energy loss to the medium can enhance the thermal partons in the vicinity of the trajectories of the semihard partons, resulting in a phenomenological structure referred to as ridge \cite{hz,hz1}.  In our present study such detailed properties of the medium are imbedded in our thermal partons, but cannot be extracted without further study of the azimuthal anisotropy.

The fact that our results 
agree well with the data at all centralities and $p_T$ attests to the reliability of our description of the minijets and their shower partons.  Despite the complexities of the geometry and nuclear medium of non-central collisions that the semihard partons must traverse to emerge at any angle, the possibility that their distribution at the medium surface can be described by a formula as simple as Eq.\ (\ref{18}) is remarkable.  Apart from the given dependence on the parton type, that formula is also universal and is crucial in determining the strengths of the shower partons that in turn are responsible for the good fits via TS, TTS, etc.\ components.  Of importance to bear in mind is the inclusion of all initiating partons that contribute to ${\hat F}_i(q,c)$, created at any point with any momentum.  In that sense it is an inclusive distribution of minijets.  The ratio $R_i(q,c)$ is analogous to the nuclear modification factor on the minijets, but is not due entirely to the momentum degradation of the hard and semihard partons. The suppression factor is also a consequence of the rapid damping of the initial distribution before momentum degradation takes effect.

A significant advance made in this study is the differentiation of the dynamical path lengths $\bar\xi_i$ for quarks and gluons. We have found in the course of our investigation that the use of an average $\bar\xi$ for all partons, as done previously, led us to a point where we could not reproduce the pion and proton data at high \ppt\ simultaneously. We were forced to recognize that the rate of energy loss for quarks and gluons are different; thus by treating $\gamma_q=\gamma_g/2$ we obtained semihard and hard parton \dis s that enabled us to fit both the pion and proton data very well. That success is not independent of the recombination formalism, since the hadrons formed in that procedure are sensitive to the parton \dis s --- pions depend more on gluons and proton more on quarks. Considering  the various roles that partons and hadrons play in our hadronization scheme, we come to the conclusion that the success in fitting the data with such accuracy would not have been possible if the partons had been treated as some generic hadron constituents. The implication is that some simple rule to convert medium to hadrons, such as parton-hadron duality, would not be a reliable way to describe hadronization when  examined closely.

Extension of this study to hyperons should be straightforward, provided that a reliable description of the relevant recombination functions can be found.  Omega production should be of particular interest because of what has been termed the ``Omega puzzle" \cite{op}.  It refers to the seemingly contradictory observation that the $\Omega$ spectrum is approximately exponential up to $p_T \sim 6$ GeV/c \cite{jad} on the one hand, while $\Omega$-triggered events have associated particles \cite{jbi} on the other.  Suppression of the strange minijets should lead to the exponential thermal behavior, but thermal partons should have no correlated partners unless they are from the ridge stimulated by non-strange minijets.

At LHC we expect both $T$ and ${\hat F}_i(q,c)$ to change, but the basic mechanism for hadronization described in Sec.\ IV does not.  Collisions at higher energy produce more partons, both thermal and shower, with increasing predominance of minijets since ${\hat F}_i(q,c)$ receives contributions from all higher initial-parton momenta $k > q$.  Based these reasonable expectations, one may regard our description as being sufficiently reliable such that it may be used as a means to discover  unexpected phenomenon, if the formalism encounters notable failure to explain some aspect of the LHC data.

\section*{Acknowledgment}

We thank R.\ Belmont for bringing  us to the attention of Ref.\ \cite{ad} and providing us with numerical values of certain data points. 
This work was supported,  in part,  by the NSFC of China under Grant No.\ 11205106  and by the U.\ S.\ Department of Energy under Grant No. DE-FG02-96ER40972.

\begin{appendix}
\section{Probability Function $P(\xi,\phi,b)$}

We summarize here the essence of the treatment of momentum degradation of semihard partons traversing ellipsoidal medium at any angle, as originally discussed in Ref.\ \cite{hy4}, but with some modifications that improve the description.  The geometrical details of determining the initial point of creation at $(x_0,y_0)$ and the exit point $(x_1,y_1)$ are not repeated here.  What is important to note is that our definition of the path length between those two points includes  weighting by the density of the medium of the static system, i.\ e.,
\begin{eqnarray}
\ell(x_0,y_0,\phi,b) = \int^{t_1(x_1,y_1)}_0 dtD(x(t),y(t)),     \label{A1}
\end{eqnarray}
where $t$ is not time, but a marker of the trajectory from $(x_0,y_0)$ to $(x_1,y_1)$, and $D(x,y)$ is the local density along the trajectory calculable from the nuclear thickness function.  The real system is not static, but the effects of expansion are mostly canceled in Eq.\ (\ref{A1}) because $D(x,y)$ decreases, while $t_1(x_1,y_1)$ increases, during expansion. Thus $\ell(x_0,y_0,\phi,b)$ may be regarded as being insensitive to hydrodynamical expansion, while being an effective measure of the distance in the nuclear medium that a semihard parton experiences in losing energy along  its path.

The probability of producing a hard (or semihard) parton at $(x_0,y_0)$ is proportional to $T_A(\vec s + \vec b /2)T_B(\vec s - \vec b /2)$, where $(x_0,y_0)$ are the Cartesian coordinates of $\vec s$ and $T_{A,B}(\vec s \pm \vec b /2)$ are the thickness functions of nuclei $A$ and $B$.  We normalize it and call it $Q(x_0,y_0,b)$ so that its integral over all $(x_0,y_0)$ is unity.  Now, we insert a $\delta$-function in the integrand and define
\begin{eqnarray}
P(\xi,\phi,b) = \int dx_0dy_0Q(x_0,y_0,b)\delta (\xi-\gamma \ell (x_0,y_0,\phi,b))     \label{A2}
\end{eqnarray}
This is the probability for the emerging parton to have had a dynamical path length $\xi$ in the medium originated from any point inside and exiting at an angle $\phi$.  The parameter $\gamma$ contains all the uncalculable effect of energy loss during the passage through the medium so that $\xi$ becomes a measure of degradation that accounts for both the geometrical length $\ell$ and the degree of degradation per unit length.  It is found in Ref.\ \cite{hy4} that a value of
\begin{eqnarray}
\gamma = 0.11     \label{A3}
\end{eqnarray}
can satisfactorily reproduce the data on $R^{\pi}_{AA}(p_T,\phi,b)$ for $p_T$ in the range 4 to 8 GeV/c.

We now improve upon the above description by recognizing that quark and gluon lose energy at different rates as they propagate through the medium. While Eq.\ (\ref{A3}) has been sufficient as an average parameter to characterize momentum degradation of partons that lead to the calculation of pion production \cite{hy4}, our interest in this article to study both meson and baryon production leads us to the necessity of distinguishing the $\gamma$ parameters for quark and gluons, whose hadronization by recombination depends on their momenta differently. To that end we use $\gamma_q$ and $\gamma_g$ for quarks and gluons, respectively, and expect them to straddle the average value $\gamma=0.11$ on its two sides. It is generally understood that gluons lose about twice as much energy as quarks \cite{pq, pq1}, so we expect $\gamma_q=0.07$ and $\gamma_g=0.14$ whose average is $\approx 0.11$ with a tilt toward gluon.

\section{Parametrization of $\bar \xi(\phi,c)$ and $\psi(z)$}

The functions $\bar \xi(\phi,b)$ and $\psi(z)$ have been studied in Ref.\ \cite{hy4} with $\gamma=0.11$. In this Appendix we proceed with the same notation with the understanding that it is only a multiplicative change from $\gamma$ to  $\gamma_i$ and $\bar\xi$ to $\bar\xi_i$. Here we give simple analytic forms for them.  Since experimental data are usually presented in terms of centrality, instead of impact parameter, we replace $b$ by $c$ that denotes the average percentile centrality, where, for example, $c=0.05$ implies 0-10\%. The relationship between $b$ and $c$ is tabulated in Ref.\ \cite{bia}.

In Fig.\ 1 of \cite{hy4} is shown a plot of $\bar \xi(\phi,c)$ vs $\phi$ for six values of $c$ ranging from 0.05 to 0.55.  An analytic approximation of that
 $\phi$ dependence can be obtained by fitting the curves in the figure with the conditions that $\bar\xi(\phi,1)=0$ and that $\bar\xi(\phi,0)$ is independent of $\phi$. We obtain the following parametrization
\begin{eqnarray}
\bar \xi(\phi,c) = 0.655 [1-c-0.32 \cos \phi \sin(c^{0.71} \pi)]  .  \label{B1}
\end{eqnarray}
It represents very well the calculated result on $\bar \xi(\phi,c)$ based on Eqs.\ (\ref{10}), (\ref{A1}) and (\ref{A3}).

For the scaling function $\psi(z)$ explicit analytic forms have been given in \cite{hy4} already.  Because the results are very insensitive to $\phi$ and $c$, we reduce them here to the simplest expression that can well represent them
\begin{eqnarray}
\psi(z) = (z/2.4)^{1/2}(1-z/2.4)/0.64.     \label{B2}
\end{eqnarray}
This function vanishes at $z=0$ and 2.4 and peaks at $z=0.8$.  It means that given $\phi$ and $c$ the most probable $\xi$ is less than the average $\bar \xi$; that is, the geometrical and nuclear complications of the collision process result in a net preference for shorter path length, independent of the degree of dynamical effect due to energy loss.

\section{Shower Parton Distributions $S^j_i$}

The SPDs are derived from the fragmentation functions (FF), $D(x)$, by treating the hadronization part of the problem by recombination, i.\ e., two shower partons in a jet recombine to form a pion.  It is well recognized that perturbative QCD treats only the $Q^2$ evolution of $D(x,Q)$ at high virtuality $Q^2$, but not how hadrons are formed.  What are needed for TS or TTS recombination in heavy-ion collisions are shower partons at intermediate $Q$.  We therefore apply the formalism of Sec.\ 2 to the FF and write for meson production \cite{hy2}
\begin{eqnarray}
xD^M_i(x) = \int {dx_1\over x_1} {dx_2\over x_2} \left\{S^j_i(x_1), S^{j'}_i\left({x_2\over 1-x_1}\right)\right\}R^M_{jj'}(x_1,x_2,x)     \label{C1}
\end{eqnarray}
where \{\ ,\ \} is a symmetrization process of the momentum fractions $x_1$ and $x_2$.  We consider only the FFs at $Q=10$ GeV/c as provided by detailed studies of the experimental data, such as in Ref.\ \cite{kkp}.  To render the determination of $S^j_i(z)$ in Eq.\ (\ref{C1}) manageable, we neglect the $Q$ dependence, and parametrize the results in the form
\begin{eqnarray}
S^j_i(z) = Az^a(1-z)^b(1+cz^d),     \label{C2}
\end{eqnarray}
where the parameters are presented in tabulated form in Ref.\ \cite{hy2}.  For $i$ and $j$ in the light-quark sector $i_q = u, d, \bar u, \bar d$, the diagonal terms of $S^j_i$ are all the same, labeled $K$, and the off-diagonal one are $L$, where $K = K_{NS} + L$ with $K_{NS}$ denoting the non-singlet valence contribution and $L$ the sea contribution.  For example, $u \rightarrow u$ has both valence and sea, but $u \rightarrow d$ has only sea.  For a gluon jet $i=g$ the symbol $G$ is used instead of $K$ or $L$.  If the produced quark is strange, $j_s = s, \bar s$, then the notation is $K_s, L_s$ and $G_s$ in place of $K, L$ and $G$, with $K_s = K_{NS} + L_s$, independent of the jet type being $i_q$ or $i_s=s, \bar s$.  We do not consider $j=g$ because we do not allow shower partons to be gluons since gluons do not hadronize directly by recombination.  For details see Ref.\ \cite{hy2}.

Since we apply $S^j_i$ to Eq.\ (\ref{26}) to calculate $S^j(p_2,c)$ for $p_2$ as low as 0.5 GeV/c, it is necessary to depart from the scaling form given in Eq.\ (\ref{C2}) and introduce a cut-off at low $p_2$.  We do that by defining (\ref{25})
\begin{eqnarray}
S^j_i(p_2,q) = S^j_i(p_2/q)\gamma_2(p_2), \qquad  \gamma_2(p_2) = 1 - e^{-(p_2/0.3)^2}.    \label{C3}
\end{eqnarray}
Correspondingly, Eq.\ (\ref{C1}) is modified, so there is also a cut-off on the scaling FF,  $D_i(x)$,
\begin{eqnarray}
D_i(p_T,q) = D_i(p_T/q)\gamma_1(p_T),  \qquad  \gamma_1(p_T) = 1 - e^{-p^2_T}.    \label{C4}
\end{eqnarray}
These cut-off cannot be rigorously derived.  We use them to round off the low-$p_T$ contributions, whose reliability is always subject to questions.

\end{appendix}



\begin{thebibliography}{99}

\bibitem{ia}
I.\ Arsene {\it et al.}  (BRAHMS Collaboration),  Nucl.\ Phys.\ A {\bf 757}, 1 (2005).

\bibitem{ba}
B.~B.~Back {\it et al.}  (PHOBOS Collaboration),  Nucl.\ Phys.\ A {\bf 757}, 28 (2005). 

\bibitem{jas}
J.\ Adams {\it et al.}  (STAR Collaboration),  Nucl.\ Phys.\ A {\bf 757}, 102 (2005).

\bibitem{ka}
K.\ Adcox {\it et al.}  (PHENIX Collaboration),  Nucl.\ Phys.\ A {\bf 757}, 184 (2005). 

\bibitem{tt}
T.\ A.\ Trainor, arXiv: 1303.4774.

\bibitem{mn}
B.\ M\"{u}ller and J.\ L.\ Nayato, Ann.\ Rev.\ Nucl.\ Part.\ Sci.\ {\bf 56}, 93 (2006).

\bibitem{vh}
P.\ F.\ Kolb and U.\ Heinz, in {\it Quark-Gluon Plasma 3}, edited by R.\ C.\ Hwa and X.-N.\ Wang (World Scientific, Singapore, 2004), p.\ 634.

\bibitem{tat}
D.\ A.\ Teaney,  in {\it Quark-Gluon Plasma 4}, edited by R.\ C.\ Hwa and X.-N.\ Wang (World Scientific, Singapore, 2010), p.\ 207.

\bibitem{hs}
U.\ Heinz and R.\ Snellings, Ann.\ Rev.\ Nucl.\ Part.\ Sci.\ {\bf 63}, xx (2013).

\bibitem{pq}
M.\ Gyulassy and X.-N.\ Wang, Nucl.\ Phys.\ B {\bf 420}, 583 (1994);  X.-N.\ Wang, M.\ Gyulassy and M.\ Plumer, Phys.\ Rev.\ D {\bf 51}, 3436 (1995).

\bibitem{pq1}
R.\ Baier, Y.\ L.\ Dokshitzer, A.\ H.\ Mueller, S.\ Peigne and D.\ Schiff, Nucl.\ Phys.\ B {\bf 
483}, 291 (1997); {\bf 484}, 265 (1997).

\bibitem{pq2}
X.-F.\ Guo and X.-N.\ Wang, Phys.\ Rev.\ Lett.\ {\bf 85}, 3591 (2000); X.-N.\ Wang and X.-F.\ Guo, Nucl.\ Phys.\ A {\bf 696}, 788 (2001).

\bibitem{pq3}
P.\ B.\ Arnold, G.\ D.\ Moore and L.\ G.\ Yaffe, JHEP {\bf 0011}, 001 (2000); {\bf 0112}, 009 (2001); {\bf 0206}, 030 (2002).

\bibitem{hy}
R.\ C.\ Hwa and C.\ B.\ Yang, Phys.\ Rev.\ C {\bf 67}, 034902 (2003).

\bibitem{vg}
V.\ Greco, C.\ M.\ Ko, and P.\ L\'{e}vai, Phys.\ Rev.\ Lett.\ {\bf 90}, 202302 (2003); Phys.\ Rev. C {\bf 68}, 034904 (2003).

\bibitem{rf}
R.\ J.\ Fries, B.\ M\"{u}ller, C.\ Nonaka, and S.\ A.\ Bass, Phys. Rev.\ Lett.\ {\bf 90}, 202303 (2003); Phys.\ Rev.\ C {\bf 68}, 044902 (2003).

\bibitem{hy1}
R.\ C.\ Hwa and C.\ B.\ Yang, Phys.\ Rev.\ C {\bf 70}, 024905 (2004).

\bibitem{rf1}
R.\ J.\ Fries, POS CERP2010, 008 (2010); arXiv:1102.5723.

\bibitem{hwa}
R.\ C.\ Hwa, in {\it Quark-Gluon Plasma 4}, edited by R.\ C.\ Hwa and X.-N.\ Wang (World Scientific, Singapore, 2010), p.\ 267.

\bibitem{cgc}
A.\ Dumitru, F.\ Gelis, L.\ McLerran, and R.\ Venugopalan, Nucl.\ Phys.\ {\bf A810}, 91 (2008).

\bibitem{cgc1}
F.\ Gelis, T.\ Lappi, and R.\ Venugopalan, Phys.\ Rev.\ D {\bf 79}, 094017 (2009).

\bibitem{cgc2}
K.\ Dusling, F.\ Gelis, T.\ Lappi, and R.\ Venugopalan, Nucl.\ Phys.\ {\bf A836}, 159 (2010).

\bibitem{cgc3}
F.\ Gelis, E.\ Iancu, J.\ Jalilian-Marian, and R.\ Venugopalan, Ann.\ Rev.\ Nucl.\ Part.\ Sci.\ {\bf 60}, 463(2010).

\bibitem{hz}
R.\ C.\ Hwa and L.\ Zhu, Phys.\ Rev.\ C {\bf 81}, 034904 (2010).

\bibitem{hz1}
R.\ C.\ Hwa and L.\ Zhu, Phys.\ Rev.\ C {\bf 86}, 024901 (2012).

\bibitem{hz2}
R.\ C.\ Hwa and L.\ Zhu, Phys.\ Rev.\ C {\bf 84}, 064914 (2011).

\bibitem{hwa1}
R.\ C.\ Hwa,  Phys.\ Rev.\ D {\bf 22}, 1593 (1980).

\bibitem{hy2}
R.\ C.\ Hwa and C.\ B.\ Yang, Phys.\ Rev.\ C {\bf 70}, 024904 (2004).

\bibitem{hy3}
R.\ C.\ Hwa and C.\ B.\ Yang, Phys.\ Rev.\ C {\bf 73}, 064904 (2006).

\bibitem{hy4}
R.\ C.\ Hwa and C.\ B.\ Yang, Phys.\ Rev.\ C {\bf 81}, 024908 (2010).

\bibitem{sa}
S.\ Afanasier {\it et al}., (PHENIX Collaboration), Phys.\ Rev.\ C {\bf 80}, 054907 (2009).

\bibitem{hy5}
R.\ C.\ Hwa and C.\ B.\ Yang, Phys.\ Rev.\ C {\bf 79}, 044908 (2009).

\bibitem{ds}
D.\ K.\ Srivastava, C.\ Gale, and R.\ J.\ Fries, Phys. Rev.\ C {\bf 67}, 034903 (2003).


\bibitem{fms}
R.\ J.\ Fries, B.\ Mueller and D.\ K.\ Srivastava, Phys.\ Rev.\ C {\bf 72}, 041902 (2005).

\bibitem{33a}
S.\ S.\ Adler {\it et al.}  (PHENIX Collaboration), Phys.\ Rev.\ C {\bf 69}, 034909 (2004).

\bibitem{ct}
C.\ Tsallis, Stat.\ Phys.\ {\bf 52}, 479 (1988).

\bibitem{kkp}
B.\ A.\ Kniehl, G.\ Kramer and B.\ P\"{o}tter, Nucl.\ Phys.\ B{\bf 597}, 337 (2001).

\bibitem{hy6}
R.\ C.\ Hwa and C.\ B.\ Yang, Phys.\ Rev.\ C {\bf 66}, 025025 (2002).

\bibitem{hy7}
R.\ C.\ Hwa and C.\ B.\ Yang, Phys.\ Rev.\ C {\bf 66}, 025024 (2002).

\bibitem{ssa}
S.\ S.\ Adler {\it et al.}  (PHENIX Collaboration), Phys.\ Rev.\ Lett.\ {\bf 91}, 072301 (2003).

\bibitem{aa}
A.\ Adare {\it et al.}  (PHENIX Collaboration), Phys.\ Rev.\ Lett.\ {\bf 101}, 232301 (2008); Phys.\ Rev.\ C {\bf 87}, 034911 (2013).

\bibitem{ja}
J.\ Adams {\it et al}., (STAR Collaboration), Phys.\ Rev.\ C {\bf 97}, 152301 (2006).

\bibitem{ga}
A.\ Agakishiev {\it et al}., (STAR Collaboration), Phys.\ Rev.\ Lett.\ {\bf 108}, 072302 (2012).

\bibitem{ad}
A.\ Adare {\it et al.}  (PHENIX Collaboration), arXiv: 1304.3410 (2013).

\bibitem{rhz}
R.\ C.\ Hwa and L.\ Zhu, arXiv: 1101.1334 (unpublished), a revised version of which is Ref.\ \cite{hz1}.

\bibitem{ph}
P.\ Huovinen, in {\it Quark-Gluon Plasma 3}, edited by R.\ C.\ Hwa and X.\-N.\ Wang (World Scientific, Singapore, 2004), p.\ 600.

\bibitem{hy8}
R.\ C.\ Hwa and C.\ B.\ Yang, Phys.\ Rev.\ Lett.\ {\bf 97}, 042301 (2006).

\bibitem{op}
R.\ C.\ Hwa, J.\ Phys.\ G: Nucl.\ Part.\ Phys.\ {\bf 34}, S789 (2007).

\bibitem{jad}
J.\ Adams {\it et al}., (STAR Collaboration), Phys.\ Rev.\ Lett.\ {\bf 92}, 182301 (2004).

\bibitem{jbi}
J.\ Bielcikova (for STAR Collaboration), J.\ Phys.\ G: Nucl.\ Part.\ Phys. {\bf 34}, S929 (2007).

\bibitem{bia}
B.\ I.\ Abelev {\it et al}., (STAR Collaboration), Phys.\ Rev.\ C {\bf 79}, 034909 (2009).



\end{thebibliography}
\end{document}